\documentclass[11pt,aps,amsmath,amssymb,nofootinbib,notitlepage,longbibliography]{revtex4-1}
\usepackage{amsmath,amssymb}
\usepackage{slashed}
\usepackage{graphicx}
\usepackage{bm}
\usepackage[top=1.0in,bottom=1.0in,left=1.0in,right=1.0in]{geometry}
\usepackage[colorlinks,linkcolor=blue,citecolor=blue]{hyperref}

\newcommand{\be}{\begin{equation}}
\newcommand{\ee}{\end{equation}}
\newcommand{\bea}{\begin{eqnarray}}
\newcommand{\eea}{\end{eqnarray}}
\newcommand{\ba}{\begin{eqnarray}}
\newcommand{\ea}{\end{eqnarray}}

\begin{document}

\title{QCD Sum Rules at High Temperature}

\author{Thors Hans Hansson}
\email{hansson@fysik.su.se}
\affiliation{Physics Department, University of Stockholm, Vanadisvagen 9 S-11346 Stockholm, Sweden}

\author{Ismail Zahed}
\email{ismail.zahed@stonybrook.edu}
\affiliation{Center for Nuclear Theory, Department of Physics and Astronomy, Stony Brook University, Stony Brook, New York 11794--3800, USA}

\begin{abstract}
We generalize the sum rule approach to investigate the nonperturbative structure of QCD at high temperature. Salient features of the QCD phase
above $T_c$ are discussed,  and included in the form of power corrections or condensate insertions, in an operator product expansion of gauge invariant
correlators. It is shown that for a plausible choice of condensates, QCD at high temperature exhibits color singlet excitations in the vector channels, as opposed to merely
screened quarks and gluons.
\\
\\
\centerline{\it This paper was posted on the SLAC server on June 5, 1990}
\end{abstract}

\maketitle

\section{Introduction}

The aim of this paper is to merge recent ideas about the nature of the 
high temperature phase in QCD, with the techniques of QCD sum rules, 
to obtain information about real time correlation functions at finite temperature. We should emphasize that these response functions can be calculated
neither in straightforward perturbation theory (because of infrared 
divergences) nor by lattice Monte-Carlo techniques (which only provides 
equilibrium quantities at finite $T$).
 That pure Yang-Mills theories exhibit a phase transition from a low temperature confined phase to a high temperature unconfined one, is by now 
well established by extensive Monte-Carlo simulations (for a. recent review 
see e.g.~[1]). When light, dynamical quarks are included, as in QCD, the 
situation is more controversial, but it has been rather convincingly demonstrated that chiral symmetry is restored at high $T$~[1]. Although the high  $T$
phase is characterized by a coupling constant $g(T)$ which is small because 
of asymptotic freedom, perturbation theory is still plagued by infrared divergences. The simplest way to understand the origin of these singularities 
is by noting that at high $T$ the 4-dimensional pure YM-theory reduces to 
to a 3-dimensional gauge-Higgs model, where the time component of the 
gauge potential plays the role of an adjoint Higgs field~[2,3,4]. The Higgs 
theory is super-renormalizable and confining with a non-perturbative mass 
gap proportional to the (dimensionful) coupling constant $g^2T$. In the original 4-dimensional theory this mass gap corresponds to a non-perturbatively 
generated magnetic mass. However, even if one introduces a magnetic mass
 $\sim g^2T$ into the 4-dimensional theory, the range of validity of perturbation 
theory remains limited [5,3].

 In addition to the proper gauge transformations, i.e. those periodic on 
the imaginary time interval $[0,\beta]$, the action is also invariant under a 
class of space-independent, non periodic gauge transformations of the type: 
$\Omega(x_4=i\beta, \vec x)=Z\Omega(x_4=0, \vec x)$, where Z is an element of the center 
of the gauge group. The order parameter used to establish the finite $T$
phase transition in pure YM theories is the Polyakov loop (or Wilson line), 
$\mathrm{Tr}\, W(\vec x) =\mathrm{Tr}\, {\rm P exp} i\int_0^\beta  A_4(\vec x, x_4)$, which is invariant under proper 
gauge transformation, but transforms non-trivially under the center symmetry. In the low temperature confining phase the center symmetry is
unbroken, and $\langle \mathrm{Tr}\, W\rangle_\beta=0$. Above the deconfining phase transition, the 
center symmetry is broken and the Polyakov loop develops a nonzero expectation value. 
(For a discussion of the universality classes related to phase transitions in gauge theories we refer to the review article by Svetetsky~[6].)

While lattice calculations of the Polyakov loop have clearly established  the breaking of the center symmetry above $T_c$, it is still not clear whether 
or not the (global) gauge symmetry is also broken. Since any order parameter for this symmetry necessarily vanishes on the lattice, due to Elizur's 
theorem, this issue is not so easy to settle. There are, however, several 
calculations and arguments in the literature which indicate that the global 
gauge symmetry is indeed spontaneously broken and that $A_4$ develops a 
non-zero expectation value. Since this idea will be very important in our 
derivation of the finite $T$ sum rules, we shall give a brief review of the 
present situation in the next section. Before that, however, we will explain the main ideas of our approach.
             
Ever since their introduction in the late seventies, the QCD sum rules have been important tools in QCD based 
phenomenology\footnote{For a comprehensive list of references to the early sum rule literature, see~[7].}. The basic idea 
of Shifman, Vainshtein and Zakharov  (SVZ) is to extract information about 
resonance masses, widths, and coupling strengths by connecting them via 
dispersion relations to various current-current correlation functions in the 
Euclidian region. These correlation functions are then calculated using 
a short distance operator product expansion (OPE). The leading term in 
this expansion corresponds to the usual perturbative contribution, which, 
because of asymptotic freedom, is well behaved for large enough Euclidian momenta ($Q^2$). The non-leading, or "condensate", terms are present 
because of the non-perturbative nature of the QCD ground state which 
implies that e.g. ($\langle 0|G_{\mu\nu}^aG_a^{\mu\nu}|0\rangle\neq 0$. Since the condensates are dimensionful, 
the corresponding contributions to the correlation functions are suppressed 
by powers of $Q^2$. What makes the QCD sum rules phenomenological rather 
than fundamental, is that a certain amount of arbitrariness goes into the 
procedure of matching the OPE calculation to the observable singularity 
structure in the physical region. First, it is necessary to assume a certain 
number of resonances and thresholds, and weight the sum rules so that they 
give sizable contribution to the dispersion integral. Second, since only few
terms in the OPE can be calculated some criteria of "perturbative dominance" (i.e. that the condensate terms are small) must be imposed. In 
spite of these difficulties, the QCD sum rules have been quite successful in  
describing a wide range of low $Q^2$ hadronic physics.

  It is clearly a challenge to extend the applicability of the QCD sum rules  
to systems with finite temperature and density, and there have already been 
several papers on this subject~[8,9,10,11]. The main idea has been to calculate the coefficient functions in finite $T$ perturbation theory using the 
same Feynman graphs as at zero temperature, and then write a sum rule 
for the retarded commutator of currents. Using this method, and assuming the condensate contributions to be essentially constant, Bochkarev and 
Shaposhnikov studied the temperature dependence of the $\rho$  mass. In their 
analysis the $\rho$  resonance disappeared at $T\sim  140\,MeV$, which they interpreted as a signal for the finite $T$ deconfinement phase transition. This 
conclusion was later challenged by Dosch and Narrison~[10] and also by 
Furnstahl, Hatsuda and Lee [11]. All these papers use the same basic idea, 
and the discrepancies between them is due to the use of different "sum rule 
technology". This emphasizes the point made above about the phenomenological nature of the QCD sum rules - the same basic strategy can lead to 
quite different conclusions. In all these works the sum rules were applied 
at temperatures around and below the transition temperature. While we 
will mainly consider the high temperature region, we will also comment on 
the use of the sum rules at low temperature. Similar comments can also be 
found in a recent investigation by Adami, Hatsuda and Zahed who critically 
analyse the breakdown of the sum rule approach at low temperature~[12].

  In this paper we discuss certain basic questions which are of importance 
for the understanding of the finite temperature sum rules. Recall that in the 
usual QCD sum rules the OPE provides a separation of scales in that the 
coefficient functions which are sensitive to the large external $Q^2$ are assumed 
to be calculable in perturbation theory, while the condensates, which are 
sensitive to long wavelengths ( i.e. $1 /\Lambda_{QCD}$) vacuum fluctuations, must be 
taken as input parameters. At finite temperature the picture is much more 
complicated due to the new scale $T$. Another, and as we shall see, related 
question is the very definition of the condensates at finite temperature. As  
discussed by SVZ, there is an implicit renormalization point dependence 
in the condensates, and an important assumption in using the QCD sum
rules is that this can be ignored~[13]. At temperatures close to and above the 
transition temperature, the structure of the QCD vacuum changes drastically (deconfinement, chiral symmetry restoration) and it is not clear how 
the condensates compare with those at zero temperature. It will turn out 
that the mechanism for deconfinement is quite important. Specifically, if 
we assume, as mentioned above, that the gauge symmetry is spontaneously 
broken there will be important new contributions to the sum rules from the 
"condensate" $\langle A_4^2\rangle$.

 We shall discuss these questions and offer as a tentative conclusion that 
the sum rules are not applicable around $T_c$, but that they might be at much 
higher temperatures, provided a suitable resummation is performed. We 
also argue that the sum rules can be applied at low temperature ($T/T_c\ll 1$), 
but that this will necessarily involve the determination of the $T$ dependence 
of the condensates (${\it confere}
$~[12]).

 The possibility of using modified QCD sum rules at temperatures well 
above the deconfining phase transition is quite interesting since it could 
provide a probe of the nonperturbative dynamics of the QCD plasma. In 
particular it gives us a possibility to test DeTar's conjecture concerning 
dynamical confinement in the plasma~[14]. The later part of this paper, 
where we investigate a QCD sum rule for the vector channel well above the 
critical temperature, will be devoted to this problem.

 The paper is organized as follows. In the next section we briefly summarize various results that are of relevance to the non-perturbative structure of 
high temperature QCD with a special emphasis on gauge symmetry breaking. In section 3 we discuss the nature of the OPE at finite temperature. in 
particular we show how calculating coefficient functions using finite temperature Feynman rules amounts to summing contribution from an infinite set 
of higher dimensional operators. We also discuss the role played by various 
mass scales in determining the applicability of the sum rules. In section 
4 we formulate the finite temperature sum rules for the p meson channel 
and in section 5 we discuss the results. Our conclusions are summarized in 
section 6. Some technical material related to the calculation of coefficient 
functions and Sorel transforms is collected in two appendices.

\section{The symmetry of QCD above $T_c$}

 In this section, we review the evidence for having a non vanishing expectation value for 
 $A_4^a(\vec x)$ at finite temperature. First recall that at finite 
temperature a constant $A_4^a$ has physical significance, since in general it 
cannot be transformed away by a proper gauge transformation. A more 
intuitive way to understand the same thing is to realize that a constant $A_4^a$
enters the theory just like an imaginary chemical potential. A two loop calculation of the partition function for pure SU(2) YM theory in a constant 
background color electric potential, was first performed by Anishetty~[15], 
who noticed that the free energy is minimized for a non zero value $gT$. 
Such a constant background potential improves the infrared behaviour of 
the theory, since it generates a mass for those gluons which are orthogonal 
to it in color charge space. The remaining massless gluons will however generate IR divergences in perturbation theory. Using dimensional reduction 
to a 3 dimensional gauge-Higgs model as referred to above, Dahlem later argued that the massless modes get a magnetic mass of similar magnitude~[16]. 
His argument was based on Polyakov's analysis of 3 dimensional compact 
QED~[17], i.e. the magnetic mass of the left over "photon" (diagonal gluons) 
was generated by a gas of monopoles. Later Belayev and Eletsky revised 
Anishetty's SU(2) calculation~[18], and extended it to SU(3)~[19] (the one 
loop SU(3) result was originally obtained by Weiss~[20]). Although these 
authors differ from Anishetty in the expression for the free energy, they 
find a non zero minimum at the same value. The SU(3) case is more complicated in that there are two gauge inequivalent directions in color space, 
corresponding to $A^3$ and $A^8$. The minimum of the two loop effective potential occurs in the $A^3$ direction corresponding to the symmetry breaking 
pattern SU(3)$\rightarrow$U(1)$\otimes$U(1). In a very recent paper, Enqvist and Kajantie 
extended the analysis of [19] to ${\cal O}(g^3)$ by summing the plasmon type ring 
diagrams for the diagonal (massless) gluons using a general covariant background field gauge. Their result for the effective action is explicitly gauge 
parameter dependent to ${\cal O}(g^2)$, and exhibits an instability at ${\cal O}(g^3)$~[21].

  Needless to say, one must be very careful in interpreting these results. 
Since there are still potential infrared difficulties due to the remaining 
massless modes, perturbation theory could very well lead us astray. In 
this context, Dahlem's argument, which in its essence was originally proposed by Gross, Yaffe and Pisarski~[22], is both interesting and compelling. 
This scenario has been further discussed in a recent paper by Polonyi and 
Vazquez~[23], who calculated the one loop Higgs potential in the SU(2) case, 
and showed that the criteria for Polyakov'a semiclassical monopole approximation are fulfilled.

 It is obviously a good idea to try to avoid perturbation theory altogether 
and look for gauge symmetry breaking by lattice techniques. There are two 
investigations which both claim numerical evidence for gauge symmetry 
breaking at high temperature. The first, by Polonyi and Wyld, used unitary 
gauge to extract expectation values of $A_4^3$  and $A_4^8$ from measurements of the 
(untraced) Polyakov loop~[24]. They observed a jump in $\langle A_4^3\rangle$ at $T_c$, and after 
assuming that the perturbative contributions are smooth, they concluded 
that $A_4^3$ develops a non zero expectation value.

 In a later paper, Mandula and Olgivie studied lattice propagators in 
Landau gauge. The $A_4$  propagator exhibits a clear jump at the transition 
temperature, consistent with the interpretation that the gauge propagator picks up a disconnected part, i.e. an expectation value for $A_4$. Again, 
this quantity cannot be measured directly, but assuming the disconnected 
parts to dominate over the perturbative contributions, the direction of sym- 
metry breaking can be obtained from a study of moments of $(A_4)^2$. The 
numerical results suggest alignment in the direction of $A_4^8$ rather than $A_4^3$, 
corresponding to SU(3) breaking to SU(2)$\otimes$U(1) rather than U(1)$\otimes$U(1). 
For our investigation it will not matter which of these alternatives is the 
correct one.

 Clearly much more work has to be done to clear up the situation, but 
we think there is enough circumstantial evidence in favour of gauge symmetry breaking at high temperature that this possibility warrants serious 
consideration. In this paper we do so by studying the effects on dynamical 
quarks with special emphasis on bound state formation.

\section{The Operator Product Expansion at finite T}

 For a general background to the OPE as applied to QCD sum rules we 
refer to the articles by Novikov, Shifman, Vainshtein and Zakharov~[25,26]. 
An essential point made by these authors is that in order to define the 
OPE, it is necessary to fix a renormalisation scale $\mu$. This scale is used to 
separate hard and soft momenta, the former contributing to the coefficient 
functions and the latter to the condensates. In particular, this means that 
there are perturbative contributions to the various condensates. In various 
toy models these perturbative pieces can be calculated explicitly, and they 
do not have to be small compared to the non perturbative contributions. 
For the special case of QCD, Novikov et.al. argue that the renormalisation 
point can be chosen such that $\alpha_s$, is small enough to allow a low order 
perturbative calculation of the coefficient functions, and at the same time 
having only a negligible $\mu$-dependence in the vacuum condensates (they 
estimate  $\leq 20\%$ for $\langle 0|G_{\mu\nu}^aG_{a^{\mu\nu}}|0\rangle$)~[25]. 
We shall see, that having a heat
bath also introduces perturbative contributions to the matrix elements in 
the OPE, and that these in general cannot be neglected.

  To illustrate this basic point we shall derive an OPE in a free massless 
scalar theory defined by
 \bea
 {\cal L}=\frac 12\partial_\mu\phi\partial^\mu\phi-\frac{m^2}2\phi^2
 \eea
and the usual canonical commutation relations. We consider the retarded 
commutator defined by
\bea
\Pi_R(x)=\bigg[:\phi^2(x):, :\phi^2(0):\bigg]\,\theta(x^0)
\eea
where $:...:$  denotes normal ordering with respect to the vacuum state. 
Define the retarded response function at finite temperature by
\bea
R_\beta(x)={\rm Tr}\bigg[e^{-\beta(H-F)}\Pi_R(x)\bigg]\,,
\eea
where the thermal average is carried out over all physical states with $F$
being the free energy. By successive commutations one easily derives the 
following OPE for $\Pi_R(x)$
\bea
\Pi_R(x)=&&-2i\Delta_R(x)\tilde\Delta (x)\nonumber\\
&&+\sum_{\mu_1...\mu_n}x^{\mu_1}...x^{\mu_n}\frac 1{n!}:\phi(0)\partial_{\mu_1}...\partial_{\mu_n}\phi(0):\,\,
\eea
where $\tilde\Delta(x)=\langle 0|\{\phi(x),\phi(0)\}|0\rangle=-2{\rm Im}\Delta_F(x)$. 
It can be shown (see.
e.g.~Appendix B in~[27]) that the first term correspond to the Feynman diagram in 
Fig.~1, calculated in Euclidian space, and analytically continued to 
real time according to the Baym-Kadanoff theorem. In this simple example 
we can of course easily evaluate the thermal trace of all the operator matrix 
elements in (4):
\bea
\langle :\phi(0)\partial_{\mu_1}...\partial_{\mu_n}\phi(0):\rangle_\beta=\int \frac{d^3k}{(2\pi)^3}\frac 1k n_\beta (ik_{\mu_1}...ik_{\mu_n})
\eea
where $n_\beta(k)$ is the Bose distribution function. Substituting in (4) and 
summing, we obtain after some algebra,
\bea
R_\beta(k)=\int \frac{d^4k}{(2\pi)^4} e^{ikx}(1+n_\beta(k_0))\,{\rm Im}\Delta_F(k)\,\,.
\eea
Again following [27] one can show that this whole contribution is again 
nothing but the Feynman graph in Fig.\,1, but this time calculated using the 
Matsubara rules for finite temperature field theory.

    The lesson of this simple example is that by calculating coefficient functions with finite temperature Feynman rules, one is in effect resuming 
contributions from higher order 
operators.\footnote{A remark in this respect is also made in [28].} 
Although we have only considered the most elementary example, we shall take for granted that this holds 
true in general. Specifically, for QCD we shall assume that by calculating 
coefficient functions in finite T perturbation theory, we systematically resum order $T$  (as opposed to order $gT$) perturbative effects in the matrix 
elements of higher dimensional operators.

    In the previous section, we presented the arguments for spontaneous 
breaking of the gauge symmetry in the Euclidian theory describing QCD 
at high $T$. We now discuss the effect of this on finite T real time correlation 
functions of the form
\bea
\Pi_{\mu\nu}(x)=[J_\mu(x), J_\nu(0)^2]\theta(x^0)
\eea
Here 4 is a color singlet current, so the operator product expansion of the 
commutator will only include color singlet operators. As described above,
we can now effectively sum contribution from infinitely many higher order 
operators by calculating the coefficient functions with finite $T$ graph rules in 
Euclidian space. We are thus led to consider graphs like those in Fig. 2. The 
general rule for the OPE is that when a low momentum is flowing through 
any of the internal lines, the graph is considered as a matrix element for 
the corresponding (gauge invariant) operators. In the figure such a low 
momentum insertion is denoted by a dot on the corresponding propagator.

   At zero temperature it is easy to evaluate the diagrams in Fig. 2b and 
2c by using the fixed point gauge $x_\mu A^\mu=0$, where the gauge potential can 
be expanded in the field strength $F_{\mu\nu}$, and its covariant derivatives[29]. The 
result is the usual QCD OPE used by Shifman, Vainshtein and Zakharov.

   At finite temperature the situation changes. Since we have shown that 
the OPE expression for the retarded commutator can be calculated from 
Euclidian finite $T$ field theory, we must consider the effects of the spontaneous breaking of the gauge symmetry. As already discussed in connection 
with the work of Mandula and Oligive, this breaking gives rise to a disconnected part in the $A_4$  gluon propagator. We can formally take this effect 
into account by including the operator $(A_4)^2$ in the OPE, and calculate its 
coefficient function in finite T perturbation theory. Alternatively we can 
keep a constant background (Euclidian)  $A_4$  potential in the calculation of 
the coefficient functions which is equivalent to introducing an imaginary 
chemical potential in the thermodynamical ensemble. The advantage with 
the first formulation is that the direction of the symmetry breaking never 
enters ({\it confere} the discussion in the previous section). In the second formulation it does, but to leading order we are only sensitive to ($(A_4)^2$). The 
advantage with the second formulation is that it emphasizes that nothing 
strange happens to the OPE. We shall elaborate a little bit on this quite important point. Consider QCD in the $A_0=0$  gauge, so that the counterpart 
to (3) will read
\bea
\langle ....\rangle_\beta={\rm Tr}[Pe^{-\beta(H-F)}....]
\eea
where $P$ projects on the physical states, i.e. it enforces the Gauss' law 
constraint. By a standard procedure we can rewrite this expression as an 
Euclidian theory with an $A_4$ Lagrange multiplier field that implements the 
constraint. In the resulting theory, $A_4$ enters as the fourth component of 
the gauge potential. Thus the whole effect of the dynamics of the $A_4$ field
is to implement the constraint on the states in the thermal ensemble. If the 
infrared behaviour of the resulting effective Euclidian theory is such that 
the long range fluctuations give rise to a vacuum expectation of $A_4$, this will 
affect the coefficient functions of the various operators in the OPE. Thus, 
introducing a non zero $(A_4)^2$ amounts to making an infinite resummation 
of terms in the OPE, which is necessitated by the vacuum rearrangement 
above $T_c$~\footnote{ In static gauges these effects are reexpressible in terms of the eigenvalues of the 
Polyako loop.}. It does not imply that a new operator $(A_4)^2$ has appeared in 
the OPE.

  To summarize, by calculating coefficient functions in the OPE,  using 
Euclidian finite $T$  perturbation theory, we sum an infinite number of contributions from higher dimensional operators. Also, in QCD we get new 
contributions (that superficially look like arising from a new gauge non 
invariant operator $(A_4)^2$ which again arise from a complicated resummation 
of terms.

  To stress these points we will make use of the formulation in terms of a 
background colored electric potential, i.e.we shall use the ensemble
\bea
\langle ....\rangle_\beta={\rm Tr}[P e^{iA_4^3Q^3}Pe^{-\beta(H-F)}....]
\eea
where $Q^a$  is the (global) color generator. We will arbitrarily take the 
symmetry breaking to be in the 3 direction of SU(3) color space (confere the 
discussion in section 2).

  We shall now ask under what circumstances the resumrnations implied 
by this formulation can be justified. We have to consider the following 
scales: the temperature $T$; the zero temperature confining scale $\Lambda_{QCD}\sim T_c$; 
the electric mass $gT$ (plasmon scale); the magnetic mass $g^2T$ 
(nonperturbative scale above $T_c$); the external momentum in the correlation function 
$Q^2$ and the renormalization point $\mu$.


  First consider $T<T_c$. In this case $g$  cannot be assumed to be small 
and one cannot distinguish between $T$, $gT$ and $g^2T$. Just as at $T=0$
one can look for a "window" where the ($T = 0$) perturbative contribution 
dominates the OPE. In this case not only the condensate terms, but also 
terms proportional to powers of ($T^2/Q^2$) have to be small. It is of course 
again possible to resum the ($(T^2/Q^2)^n$) terms to obtain coefficient functions 
calculated in finite $T$ perturbation theory as discussed above. For these low 
temperatures, however, such a resummation does not seem to make much 
sense since the thermal fluctuations will dominantly be at low momenta. 
As such they should not change the coefficient functions, but rather be 
included in the operator matrix elements. Thus, it is reasonable to believe 
that at  $T<T_c$  the main temperature effects are in the condensates and 
not in the coefficient functions. Hopefully, improved lattice calculations will 
enable us to actually calculate the $T$  dependence of the most important low 
dimensional condensates like $G^2$  and $\bar qq$. When the temperature is about $T_c$, 
some of the condensates are expected to vary rapidly with the temperature. 
In this range the QCD sum rule approach is likely to break down and lose 
its predictive power, (compare with reference [12]). These comments are 
relevant for the calculations presented in references [8,10] and [11].

 The discussion so far indicates that the issue of scale separation is problematic at low temperature where the range of applicability of the sum rule 
approach might be drastically constrained. Fortunately, the situation is 
much better at high temperature, or more specifically, in the regime of 
temperatures such that
\bea
T_c<g^2T<Q^2<T
\eea
The electric scale $gT$ and the renormalization point $\mu$  will be discussed later.  
First, since we are at high $T$ we can hope to get some non-perturbative information about the condensates both from lattice calculations [30,31] and 
estimates in the effective 3 dimensional gauge-Higgs model [32,33]. 
Secondly, $Q^2$ can hopefully be made small enough to really probe the meson 
like excitations in the plasma while still keeping the OPE under control 
(remember that the scale of the finite $T$ condensates is $g^2T$), i.e. we can 
hope for a "window". Third, the large ($T^2/Q^2$) power corrections must be 
resummed, but this is achieved by the methods described above. Thus we 
propose that the region specified by (10) is the one that can be studied 
using the resummed OPE described earlier in this section.

 Several comments on this suggestion, on which this paper is based, are 
in order. First, it is clear that for the hierarchy (10) to make sense, the 
temperature must be sufficiently large for $g(T)$ to be small. Since we don't 
know the relevant expansion parameter ( $g(T)$ or $g(T)/2\pi$ or ...), what 
is "sufficiently large $T$ cannot be simply answered by e.g. using the one
loop formula for the running coupling constant. Thus, for the time being 
we can only assume that (10) makes sense at temperatures that might be 
experimentally accessible. In calculating the coefficient functions, we shall  
use ordinary finite $T$ Feynman rules. We know, however, that an electric 
mass is generated at the scale $gT$. We already discussed the resummation of 
terms of order $(T^2/Q^2)^n$, but if  $Q^2$  is not much larger than $(gT)^2$, we must 
also resume the $(g^2T^2/Q^2)^n$ terms, 
by introducing an electric mass.\footnote{ A first attempt in this direction can be found in [21].} 
In the 
following we will for simplicity assume that $Q^2$ can be chosen so that this 
is not needed. Note that corrections due to magnetic mass insertions will 
be of the form $(g^4T^2/Q^2)^n$, and are thus included in the matrix elements. 
At $T=0$  the operator renormalization scale $\mu$  is chosen so as to minimize 
perturbative corrections to the condensates. At finite $T$ we would expect 
this to happen for $g^2T<\mu<Q^2$, but just like at $T = 0$, this cannot be 
proved but will be assumed.

\section{Sum rules}

 To investigate the possible occurrence of light bound states above $T_c$ we 
will construct sum rules for the thermal average of the retarded correlation 
function for the color singlet vector-isovector ($1^{--}$) current 
$J_\mu=\frac 12(\bar u\gamma_\mu u-\bar d\gamma_\mu d)$. 
The Fourier transform of this correlator reads,
\bea
\Pi_{\mu\nu}(q_0, \vec q)=i\int d^4xe^{iq\cdot x}{\rm Tr}(
Pe^{iA_4^3Q^3}e^{-\beta(H-F)}[J_\mu(x), J_\nu(0)]
\eea
Let us recall that $P$  is the projection operator on physical states, and that 
the term $e^{iA_4^3Q^3}$ enforces the spontaneous breaking of the global color 
symmetry related to the vacuum rearrangement above the critical temperature. 
Since both $H$ and $P$ commute with the global charge, we can use standard 
methods to rewrite (2) as an Euclidian path integral in a constant background 
$A_4^3$ potential. We shall use covariant gauge, and as usual the ghosts 
obey Bose statistics. At finite temperature, the retarded correlator (11) is
determined by two invariant form factors,
\bea
G_{ij}&=&\bigg(\delta_{ij}-\frac{Q_iQ_j}{\vec Q^2}\bigg)G_T+\frac{Q_iQ_j}{\vec Q^2}Q_0^2G_L \\
G_{00}&=&\vec Q^2 G_L
\eea
where $q=(iQ_0, \vec 0)$.  Both are analytic in the upper $Q^2$ plane. As a result,
the longitudinal and transverse form factors satisfy conventional dispersion 
relations. Specifically, $G_L$  obeys an unsubtracted dispersion relation of the 
type
\bea
ReG_L(Q^2)=\frac 1\pi\int_0^\infty ds\frac{{\rm Im}G_L(s)}{s+Q^2}
\eea
In the frame of the heat bath, $Q=(Q_0, \vec 0)$, the transverse form factor $G_T$ is 
related to the longitudinal form factor since $G_T\rightarrow Q^2 G_L$
 when $\vec Q^2\rightarrow 0$~[28). 
Hence, only the longitudinal form factor will be discussed throughout.

 For large space-like $Q^2$, the real part (LHS) can be calculated using the 
OPE. The imaginary part, or spectral density, in the RHS is given by poles 
and cuts in the time-like region which cannot be approached in perturbation 
theory. In the sum rule approach this part is conventionally parametrized 
by a single resonance pole and a continuum threshold. Sum rules are then 
derived for the position and residue of the resonance, and the position 
and the strength of the threshold. There are various methods proposed 
for matching the RHS and LHS in the sum rules. For light resonances 
the most common procedure is to Borel transform the expressions and 
look for stability of the difference RHS-LHS as a function of the Borel 
parameter $M^2$. Borel transformation usually improves the convergence of 
the (asymptotic) perturbative expansion (factorial suppression), and also 
further suppresses the contributions from the higher dimensional operators. 
Because of the large uncertainties involved in the present calculation, we 
have not explored the various other formulations of the sum rules that have 
been proposed.\footnote{For a rationale of the Borel method we refer to the original papers by SVZ and for a 
recent discussion of these matters in the context of finite T sum rules to [10].}

 Let $m_\rho$ and  $f_\rho$  be the the mass and residue respectively, of a {\it  possible} 
light bound state in the $\rho$ channel above $T_c$ and let  $\sqrt{s_0}$ be the threshold
energy of the continuum (in this picture we implicitly assume dynamical  
confinement, in the sense that the resonance is isolated from, rather than  
imbedded in, the continuum). Then the RHS of (14) can be parametrized  
as follows
	
\bea
{\rm Im}G_L(s)=\pi f_\rho m_\rho^2\delta(s-m_\rho^2)+\theta(s-s_0)\frac 1{8\pi}
{\rm tanh}\frac{\sqrt{s}}{4T}+\frac{2\pi T^2}{3}\delta(s)
\eea
As in the conventional sum rules we have chosen to parametrize the $q\bar q $
background, i.e. the strength of the cut, by the (temperature dependent)  
imaginary part of the perturbative value. The delta function contribution  
in (15) corresponds to soft scattering of thermal quarks off the resonance  
(thermal dissociation)~[28].

The OPE is now used to evaluate the LHS of (14) in the deep Euclidean  
regime ($Q^2\rightarrow 0$). At high temperature, however, nontrivial resummations  
are needed. The rationale for this has been described in section 3. In this  
spirit, we will evaluate the Wilson coefficients in Euclidean space using the  
Matsubara techniques and analytically continue them to Minkowski space  
using the Baym-Kadanoff prescription. Figs. 2b and 2c show the leading  
nonperturbative contributions to the causal correlator as implied by the  
OPE. In these figures, the blob can mean either ($\langle B^2\rangle$), 
($\langle E^2\rangle$) or $\langle(A_4)^2\rangle$.  
The $(A_4)^2$ insertions are due to the complex colored chemical potential  
inserted in (11) to account for the spontaneous breakdown of global gauge  
invariance. The $E^2$ and $B^2$ insertions correspond to the usual electric and  
magnetic condensates which are still present above $T_c$. These need not be  
the same as at T = 0 since the heat bath defines a preferred Lorentz frame  
which means that we have the two O(3) invariants $E^2$ and $B^2$  rather than  
the single O(4) invariant $G^2$. Higher dimensional operators are suppressed  
by powers of $Q^2$. Remember that chiral symmetry is restored above $T_c$ so,  
at least for massive quarks, there are no quark condensate insertions. The  
results for the Wilson coefficients are

\bea
{\cal C}_1&=&+\frac 1{8\pi^2}\int_0^\infty ds\bigg(1+8n_\beta\frac s{Q^2}\bigg)\frac 1{s+Q^2/4}\\
{\cal C}_{A_4^2}&=&+\frac {\alpha_s}{12\pi}\int_0^\infty ds\,2n_\beta\frac{2-3Q^2/4}{(s+Q^2/4)^3}\\
{\cal C}_{E^2}&=&-\frac {\alpha_s}{6\pi} \frac 1{Q^4}-\frac{\alpha_s}{72\pi}\frac 1{Q^2}
\int_0^\infty ds\,2n_\beta\frac{2-3Q^2/4}{(s+Q^2/4)^3}\\
{\cal C}_{B^2}&=&+\frac {\alpha_s}{36\pi}\frac 1{Q^2}\int_0^\infty ds\,2n_\beta\frac{2-3Q^2/4}{(s+Q^2/4)^3}
\eea
where $n_\beta = n(\sqrt{s}/T)$ is the Fermi distribution. A derivation of these 
coefficients is given in Appendix A using a variant of the fixed point gauge.

The Wilson coefficients for $E^2$  and $B^2$  corresponding to Fig.~2b are 
infrared divergent both at zero and finite temperature, since the momentum 
in the dressed quark line can become soft. At zero temperature Lorentz 
invariance and gauge invariance conspire to the vanishing of Fig.~2b in the 
chiral limit, but this is no longer true for massive fermions. Fortunately, 
this soft contribution can always be reabsorbed into $m\bar q q$ at the operator 
level using the equations of motion. This prescription, which is due 
to Smilga~[29], is shortly discussed in Appendix A, and will be understood 
throughout. As a result, the contribution of Fig.~2b will be set to zero even 
at finite temperature since we are all the time working in the chiral limit.

 Following the conventional recipe used for for light $Q^2$ systems at zero 
temperature, we equate the Borel transform of the RHS with that of the 
the LHS, to get the following sum rule,

\bea
f_\rho^2m_\rho^2e^{-m_\rho^2/M^2}=\frac 1{8\pi^2}R(M^2,T)
\eea
where R is given by

\bea
R(M^2, T)=&&\,\,\,\int_0^{s_0}ds\, e^{-s/M^2}\,{\rm tanh}\bigg(\frac{\sqrt s}{4T}\bigg)\nonumber\\
&&-\frac {16\pi\alpha_s}3\frac{\langle (A_4)^2\rangle}{M^2}\int_0^\infty ds\,n_\beta\bigg(\frac{\sqrt s}{2T}\bigg)\bigg(-3+\frac {2s}{M^2}\bigg)\,e^{-s/M^2}\nonumber\\
&&-\frac{8\pi\alpha_s}9\frac{\langle E^2\rangle}{M^2}\nonumber\\
&&-\frac{4\pi\alpha_s}3 \frac{\langle(E^2-2B^2)\rangle}{M^4}\int_0^\infty ds\,n_\beta\bigg(\frac{\sqrt s}{2T}\bigg)\,B(M,s)
\eea
and

\bea
B(M,s)=\frac{M^4}{s^2}\bigg(1-e^{-s/M^2}\bigg(1+\frac s{M^2}+\frac {2s^2}{M^4}\bigg)\bigg)
\eea
which satisfies $\int_0^\infty ds B=-1$.  The pertinent Borel transforms for the finite 
temperature expressions can be found in Appendix B. Note that the threshold parametrization used in (15), implies a large cancellation between the 
continuum contributions on the two sides of the sum rule.

\section{Numerical results}

 By now it should be clear that we are not in a position to make any 
firm predictions based on the sum rule (20). In addition to all the assumptions 
made in deriving it, we also have the additional difficulties related 
to the finite $T$ condensates. At  $T = 0$ the condensates can be determined 
phenomenologically while here we must rely on theoretical estimates. With 
all this in mind, we will nevertheless try to estimate our parameters using 
available information, plug everything into (20) and look for signs of above 
$T_c$  resonances. We believe, that at this point, nothing better can be done, 
and that the results below represent a possible scenario. Needless to say, 
we have not spent any time trying to optimize our fits, explore different 
forms of the sum rules, do systematic searches in parameter space etc.. 
Such efforts would be wasted on a numerical calculation which is primarily 
illustrative. For the same reasons we shall also not present any details of 
our fits, but only the basic results.

 From the lattice measurements of the free energy and pressure, it is 
possible to make an estimate of the electric and magnetic condensates above 
$T_c$.  The analysis of Adami, Hatsuda and Zahed [12] gives 
$\langle (\alpha_s/\pi)B^2\rangle\sim \langle (\alpha_s/\pi)E^2\rangle\sim (196 \pm  10\,MeV)^4$ from lattice calculations in the range 
$T_c<T<2T_c$.  Above $T_c$ the electric and magnetic condensates are about half 
their values at $T=0$. These estimates are consistent with the recent SU(2)
lattice calculation by Lee~[31].

 Numerical estimates for the Higgs condensate are unfortunately not 
available. The lattice calculations of Mandula and Olgivie were done in 
Landau gauge and thus cannot be directly used here. For lack of better, 
we shall nevertheless take $\langle(\alpha_s/\pi)(A_0)^2\rangle\sim +0.02\, GeV^2$, which is consistent 
with their results~[33].

 The LHS and RHS of the sum rule (20) can be made to match above 
$T_c$. For $T = 1.5\,T_c$  we find a Borel "window",  $0.6 \,GeV^2 < M^2 < 1.2\,GeV^2$, 
where the sum rule is saturated by a resonance with mass $m_\rho\sim 1\,GeV$ 
and strength $f_\rho\sim 0.03$  plus a threshold at $s_0\sim 1.5\, GeV$. Recall that 
the resonance strength is a measurement of the rho coupling above $T_c$, 
i.e. $g_\rho^2/4\pi=1/4\pi f_\rho\sim 2.5$.

 For the range of parameters quoted above, the vacuum contribution is 
about three times the contribution of the condensates. More specifically,
the electric and magnetic condensates contribute about 1\% of the RHS 
while the $(A_4)^2$ condensate contributes about 30\%. This shows that the 
perturbative approach is consistent. Moreover, the continuum contribution 
to the sum rule is about 20\% of the resonance contribution, which justifies 
the single resonance dominance approximation.

\section{conclusions}

  We have shown how to use present theoretical ideas about the high 
temperature phase of QCD in order to calculate finite temperature real 
time correlation functions, and extract information about light relativistic 
bound states. A first crude attempt to a numerical analysis of the sum 
rules gives a clear indication for a resonance structure in the rho channel. 
Since chiral symmetry is restored above $T_c$, we expect the same signal to 
occur in the $A_1$  channel. Other light multiplets are also expected. We 
again emphasize that the correlators we consider can be obtained neither 
in perturbation theory, because of the infrared problems above $T_c$ nor in 
present lattice calculations because of their limitation to static quantities. 
(Attempts to analytically continue numerical results from high temperature 
lattice results are questionable.)
  Our results indicate that the high temperature phase of QCD has excitations which are reminiscent of those at zero temperature. The constituents 
of the plasma are not just screened quarks and gluons but also propagating 
color singlet excitations as first conjectured by DeTar[14]. Such light resonances, if confirmed, might have important phenomenological implications.
  Our work can be extended in several ways. First, direct lattice calculations of the lowest dimensional condensates are needed to put any numerical 
analysis of the sum rules on a firm basis. Second, more systematic investigations of the resonance parameters can be carried out for a wide range 
of temperatures above $T_c$, to study the behavior of the expected light chiral 
multiplets. Third, the analysis can be extended to baryons.

\section{Acknowledgements}

 We would like to thank Chris Adami for help in the numerical calcu- 
lations. T.H.H. would like to thank Janos Polonyi and Kejo Kajantie, and 
I.Z. Sudhir Nadkarni for interesting discussions.

\appendix

\section{}

  In this Appendix we outline the calculation of the Wilson coefficients 
(16)-(19) given in section 4. At finite temperature, and in the spontaneously 
broken phase, it is convenient to use the following gauge,
\bea
\partial_0A_0=0\qquad{\rm and}\qquad \vec x\cdot\vec A=0
\eea
which is a combination of the static gauge and the 0(3) version of the 
Fock-Schwinger, or fixed point gauge. The calculation is performed in Minkowski 
space and the final result converted to Euclidian (note $\langle A_4^2\rangle=-\langle A_0^2\rangle >0$).
It can be easily checked that the gauge conditions (A.1) are both reachable 
and complete, and imply the following relations,
\bea
A_0(t, \vec x)&=&A_0(\vec x)\nonumber\\
A_j(t, \vec x)&=&\int_0^1d\alpha\,x^i\,A_{ij}(t, \alpha\vec x)
\eea
The second of these equations shows that the vector potential can be 
expressed solely in terms of the magnetic field and its covariant derivatives, i.e.
\bea
A_j(t, \vec x)=\sum_0^\infty \frac 1{k!}(k+2)\,x^ix^{j_1}....x^{j_k}\nabla_{j_1}...\nabla_{j_k}\,A_{ij}(t, \vec 0)
\eea
Where $A_{ij}=A_{ij}^aT^a$  is the (matrix-valued) colormagnetic field tensor. In
this gauge the momentum space fermion propagator reads,
\bea
S(q)=\frac{\slashed{q}}{q^6}-\frac{\slashed q \gamma^0\slashed q}{q^4}gA_0-\frac{\slashed q}{q^6}q^jq^0g D_jA_0-\frac 1{4q^2}(\sigma^{ij}\slashed q+\slashed q\sigma^{ij})gA_{ij}+\frac{\slashed q\gamma^0\slashed q\gamma^0\slashed q}{q^6}(gA_0)^2+....\nonumber\\
\eea
The ... stands for higher dimensional contributions. By relating the 
background potential to the (Euclidian space) "condensate" according to
\bea
A_0^aA_0^b=-\frac 18\delta^{ab}\langle A_4^2\rangle
\eea
a straightforward calculation yields the following contribution to the trace 
of the polarization tensor from the diagrams 2b and 2c:
   \bea
  \Pi_{\mu\mu}^{(2b)}&=&4g^2\langle A_4^2\rangle (-3q_0 I_{21}^0+4q_0 I_{31}^{000}+2I_{21}^{00}-I_{11})\\
  \Pi_{\mu\mu}^{(2c)}&=&2g^2\langle A_4^2\rangle (2q_0^2I_{22}^{00}-q_0^2I_{12}+2q_0I_{12}^{0}+I_{02})
  \eea
In both cases we used the following notation
\bea
I_{mn}^{\mu_1...\mu_j}=\sum_\beta \frac{k^{\mu_1}...k^{\mu_j}}{k^{2m}(k+q)^{2n}}
\eea
The sum in (A.8) is over discrete energies $k_0=2\pi/\beta$  and continuous 
momenta $\vec k$. Some useful properties of these integrals can be found in reference 
[34]. Relating $G_L$  to $\Pi_\mu^\mu$  we obtain the coefficient function (16 -19) in the 
text. An alternative way to calculate the coefficient function for the $A_4^2$ insertion, is to use Feynman rules corresponding to an (imaginary) chemical 
potential.

  The magnetic contributions to the trace of the response function is 
shown in Fig. 2b and 2c, and follows from the magnetic insertion appearing 
in the fermion propagator (A.4). Specifically, the contribution of Figure 2b 
and 2c are
\bea
  \Pi_{\mu\mu}^{(2b)}&=&\frac{32 g^2}3\langle B^2\rangle (2q_0I_{31}^0+I_{31}^{00}+q_0^2 I_{41}^{00}-I_{40}^{00}) \\
    \Pi_{\mu\mu}^{(2c)}&=&-\frac{g^2}3\langle B^2\rangle (-2I_{12}-q_0^2I_{22}+4q_0I_{22}^0+4I_{22}^{00}) 
\eea
The contribution to (A.9) diverges in the infrared at zero temperature. The 
reason for this is rather simple. Each magnetic field insertion brings two 
powers of $q^2$ in the denominator turning the process divergent in the infrared. 
As discussed by Smilga, this process can be regarded as a condensate effect and can be reabsorbed into $m\bar qq$ at zero temperature using the 
equations of motion. We will assume the same to hold true at finite
 temperature. Since the fermion condensate vanishes above $T_c$. The contribution 
of Fig.~2b will be set equal to zero.

  At this stage, we should point out that if we were to use the operator 
formalism rather than the background field method, and include the short
distance effects only in the coefficient functions, the infrared problem we 
encountered would never have occurred. That it did, shows that spurious 
effects can occur when coefficient functions are calculated using the background field method. Similar arguments apply to the electric contribution 
to the trace of the polarization function. Combining these results with 
(A.6 - A.7) and ignoring (A.9) we have\footnote{The terms involving $E^2$ were calculated using the ordinary fixed point gauge. The 
terms involving $B^2$ were calculated using the ordinary fixed point gauge and the gauge 
(A.1). For the latter both gauges lead the same answer as expected.}
\bea
\Pi_{\mu\mu}=&&-4g^2\langle E^2\rangle(2I_{12}-q_0^2I_{22})\nonumber\\
&&+\frac {4g^2}3\langle E^2+B^2\rangle (2I_{12}-q_0^2I_{22}+4q_0I_{22}^0+4I_{22}^{00})
\eea
The asymmetry between the electric and magnetic contributions to $\Pi_{\mu\nu}$ 
is due to the breaking of Lorentz invariance introduced by the heat bath. 
Indeed, if we denote by $n^\mu=(1,\vec 0)$ the rest frame of the heat bath, we can 
decompose the product of two field strengths as
\bea
G_{\mu\nu}^aG^{b\alpha\beta}=-\frac{\delta^{ab}}{24}(g_\mu^\alpha g_\nu^\beta-g_\nu^\alpha g_\mu^\beta) E^2+\frac {\delta^{ab}}{24}
\epsilon_{\mu\nu\rho\sigma}\epsilon^{\alpha\beta\rho\tau} n^\sigma n_\tau (E^2+B^2)
\eea
It is rather clear that at zero temperature only the electric part contributes 
since the vacuum is Lorentz invariant (with $\langle E^2\rangle =-\langle B^2\rangle$ ). At finite 
temperature this h no longer true. Notice that the magnetic effects are 
frame dependent.

     The summation in the $I$  integrals can be carried out by separating the 
zero and finite temperature contributions. The zero temperature part can 
be calculated using conventional dimensional regularization and the results 
are
\bea
&&2I_{12}-q_0^2I_{22}=\frac{i}{8\pi^2}\frac 1{q_0^2}\nonumber\\
&&2I_{12}-q_0^2I_{22}+4q_0I_{22}^0+4I_{22}^{00}=0
\eea
 The Lorentz breaking term in (A.11) follows the decomposition (A.12) and 
 vanishes at zero temperature as it should. At finite temperature both terms in (A.11) contribute. Using
\bea
2I_{12}-q_0^2I_{22}=4q_0I_{22}^0+4I_{22}^{00}=
\frac 2{\pi^2}\int dx x n_\beta \frac{4x^2+3q_0^2}{(4x^2-q_0^2)^3}
\eea
in (A.11) and substituting  $q_0\rightarrow -iQ_0$  yield the quoted results for the coefficient functions.

\section{}

In this appendix we derive two useful expressions for the Borel transforms of inverse powers of polynomials needed in deriving (21). If we denote 
by $L_M$ the Borel transformation than
\bea
L_M\bigg(\frac 1{(s+Q^2)^d}\bigg)&\equiv& {\lim_{n\to\infty}}\frac{(Q^2)^n}{(n-1)!}\bigg(-\frac d{dQ^2}\bigg)^n\bigg(\frac 1{s+Q^2}\bigg)^d\nonumber\\
&=&{\lim_{n\to\infty}}\frac{d(d+1)...(d+n-1)}{(n-1)!}\bigg(\frac 1{Q^2}\bigg)^d\bigg(\frac 1{1+s/(nM^2)}\bigg)^{d+n}\nonumber\\
&=&\frac 1{(d-1)!}\bigg(\frac 1{M^2}\bigg)^de^{-s/M^2}
\eea
where $Q^2=nM^2$ and the last equality follows from Stirling's approximation 
in the limit of large $n$. We also have
\bea
L_M\bigg(\frac 1{Q^2(s+Q^2)^d}\bigg)&=&\frac 1s L_M\bigg(\frac 1{Q^2(s+Q^2)^{d-1}}\bigg)-\frac 1sL_M\bigg(\frac 1{(s+Q^2)^d}\bigg)\nonumber\\
&=&\frac 1{s^d}\frac 1{M^2}-\frac 1{M^{2d+2}}\sum_{k=1}^d\frac 1{(d-k)!}\bigg(\frac {M^2}s\bigg)^ke^{-s/M^2}
\eea
The last equality follows from (37) by successive iterations. Using (B.1) 
and (B.2) it is straightforward to derive the Borel transforms of the LHS of 
(14) from the expressions (16)-(19) for the Wilson coefficients. The Borel 
transform of the RHS is standard.

\newpage
\newcounter{mycount}
\newenvironment{mylist}
{ \begin{list}{ [\arabic{mycount}] }{ \usecounter{mycount} }  }{ \end{list} }

           {\bf References}
           
           \vskip 5mm 
           
\begin{mylist}

\item  F. Karsch, In Simulating the Quark- Gluon Plasma on the Lattice, \\ CERN preprint, CERN-TH-5498, 1989.

\item T. Applequist and J. Carrazzone, Phys. Rev., D 11 (1975) 2856. 
 
\item T. Applequist and R.D. Pisarski, Phys. Rev., D23 (1981) 2305. 
           
\item  S. Nadkarni, Phys. Rev., D27 (1989) 917.
           
\item A.D. Linde, Phys. Lett., 96B (1980) 289.
           
\item B. Svetetsky, Phys. Rep., 132 (1986) 1.
           
\item E. Shuryak, In The QCD Vacuum, Hadrons and the Superdense Matter, World Scientific, 1988.
            
\item  A.I. Bochkarev and M.E. Shaposhnikov, Phys. Lett., 145B (1984)276.   
           
\item  A.I. Bochkarev and M.E. Shaposhnikov, Nucl. Phys., B268  (1986) 220.
            
\item  H.G. Dosch and S. Narison, Phys. Lett., B2O3 (1988) 155.
           
\item  R.G. Furnstal, T. Hatsuda, and S.H. Lee, Applications of QCD Sum Rules at Finite Temperature,  University of Maryland preprint, UMPP-90-031, 1989.
         
\item C. Adami, T. Hatsuda, and I. Zahed, QCD Sum Rules at Low Temperature, \\ Stony Brook preprint, SUNY-NTC-89-48, 1989.
            
\item V.A. Novikov, M.A. Shifman, A.I. Vainshtain, and V.I. Zakharov, \\ Nucl. Phys., B237 (1984) 525.
            
\item  C. DeTar, Phys. Rev., D24 (1981) 475.
           
\item R. Anishetty, J. Phys. G, Nucl. Phys. (1984)10.
           
\item  K.J. Dahlem, Zeit. Phys., C 29 (1985) 553. 
           
\item  A.M. Polyakov, Nucl. Phys., B 120 (1977) 429.
\item V.M. Belyaev and V.L. Eletsky, ITEP preprint, ITEP-69, 1989. 

\item V.M. Belyaev and V.L. Eletsky, Zeit. Phys., C46 (1990) 355.

\item  N. Weiss, Phys. Rev., D 24 (1981) 475.

\item  K. Enqvist and K. Kajantie, University of Helsinki preprint, HU-TFT-90-9, 1990.
  
\item D.J. Gross, R.P. Pisarski, and L.G. Yaffe, Rev. Mod. Phys., 53 (1981) 43.
  
\item  J. Polonyi and S. Vazquez, In The Higgs Phase of QCD, MIT preprint,  1990.
  
\item  J. Polonyi and B. Wyid, In Gauge Symmetry and Compactification, \\ University of Urbana preprint, TH-85-23, 1985.
  
\item  M.A. Shifman, A.I. Vainshtain, and V.I. Zakharov, Nucl. Phys., B147  (1979) 385.
  
\item  V.A. Novikov, M.A. Shifman, A.I. Vainshtain, and V.I. Zakharov,  \\ Fort. der Phys., B32 (1984) 585.
  
\item  M. Carrington, T.H. Hansson, H. Yamagishi, and I. Zahed, Ann. Phys., 190 (1989) 373.
  
\item  A.I. Bochkarev and M.E. Shaposhnikov, Spectrum of the Hot Hadronic Matter and Finite Temperature QCD Sum Rules,  Moscow Academy of Sciences, P-0391, 1985.
  
\item  A. Smilga, Soy. Jour. Phys., 35 (1982) 271.

\item A. Di Giacomo and C.C. Rossi, Phys. Lett., 100B (1981) 481. 

\item S.H. Lee, Phys. Rev., 40 (1989) 2484.

\item S. Nadkarni, Phys. Rev. Lett., 60 (1988)491.

\item  J. Mandula and M. Olgivie, Phys. Lett., 201B (1988) 117. 

\item  T.H. Hansson and I. Zahed, Nucl. Phys., 292 (1987) 725.

\end{mylist}

\newpage

          {\bf Figure Caption}
          
          \vskip 1cm

         Fig. 1. Vacuum contribution to the retarted current-current correlator for 
         free scalars as defined in Eq. (2).

         Fig. 2. (a) Vacuum contribution to the vector current; (b) and (c) Soft 
         gluon contributions to the vector current.


\begin{figure}[h!]
	\begin{center}
		\includegraphics[width=16cm]{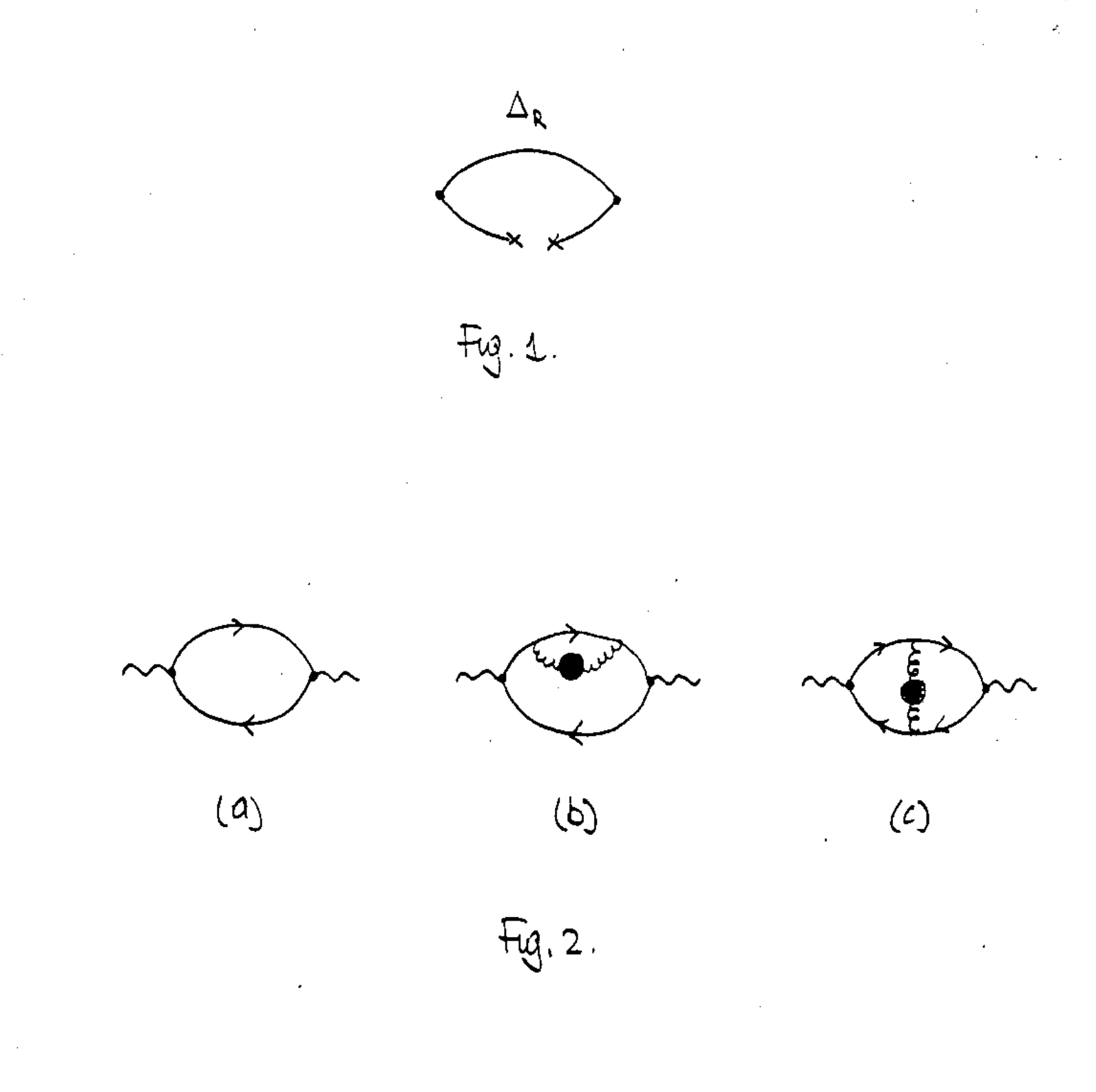}
		\caption{}
		\label{fig}
	\end{center}
\end{figure}


\end{document}